# Collective Construction of 2D Block Structures with Holes

Zachary Fitzsimmons[1] and Robin Flatland[2]

[1] Siena College, Loudonville, USA `zm19fitz@siena.edu`
[2] Siena College, Loudonville, USA `flatland@siena.edu`

**Abstract.** In this paper we present algorithms for collective construction systems in which a large number of autonomous mobile robots transport modular building elements to construct a desired structure. We focus on building block structures subject to some physical constraints that restrict the order in which the blocks may be attached to the structure. Specifically, we determine a partial ordering on the blocks such that if they are attached in accordance with this ordering, then (*i*) the structure is a single, connected piece at all intermediate stages of construction, and (*ii*) no block is attached between two other previously attached blocks, since such a space is too narrow for a robot to maneuver a block into it. Previous work has consider this problem for building 2D structures without holes. Here we extend this work to 2D structures with holes. We accomplish this by modeling the problem as a graph orientation problem and describe an $O(n^2)$ algorithm for solving it. We also describe how this partial ordering may be used in a distributed fashion by the robots to coordinate their actions during the building process.

## 1 Introduction

In this paper we present algorithms for collective construction systems in which a large number of autonomous mobile robots transport modular building blocks to construct a desired structure. Also known as swarm construction, such systems are inspired by the decentralized construction methods of ants and bees in which global structure emerges from the efforts of individual insects following seemingly simple rules and using only local environmental cues. By using many identical robots executing a common set of rules, such systems can be robust to failure since damaged robots are easily replaced. This makes them useful in uncertain and inhospitable environments such as extraterrestrial and underwater construction sites. In some variations of collective construction, the modular building blocks are not completely inert. They may, for example, have a small readable/writable memory in which the robots can write information to help guide subsequent construction.

Collective construction is related to self-reconfiguring modular robots and self assembly systems. Self-reconfiguring modular robots are composed of many simple robotic units attached to one another to form a structure. The robots have some computational power and the capacity to perform simple motion such as expanding and contracting its sides [8] or rotating [9], [10], and they can attach/detach to/from adjacent units. When the units coordinate their actions, they move relative to one another and the overall structure changes shape to fit the task or environment at hand. See for example [3] for a survey and [4], [6], [5] for recent algorithmic advances. In self-assembly systems, the units are DNA "tiles" and their operations are determined by the type and order of "glues" located on their boundaries. When tiles of particular types are mixed together, they bond according to their glues to form a structure. See for example [7]. Unlike self-reconfiguring robots and self-assembly systems, collective construction separates the passive, structural building elements from the active, functional units that combine these elements, which seems a natural separation when building larger scale, static structures.

Here we focus on a collective construction system introduced by Werfel and Nagpal [1]. In their system, the structural elements are unit cubes (blocks) that attach to each other face-to-face. Each robot is capable of carrying a block and depositing it in a desired location, at which point the block attaches to any neighboring blocks. In [1], they consider the problem of programming the robots to build connected, 2D block structures without holes. Here 2D means the structure consists of a single layer of blocks, with no block stacked on top of another. Since the blocks attach face-to-face with the faces aligned, the structures to be built can be represented by a set of marked cells on a unit grid, such as shown in Figure 1a.

A physical constraint that plays into their algorithm design is that they disallow sliding a block into a space exactly one block wide between two other blocks, since such a space is too narrow for a robot to maneuver a block into it. See for example the unoccupied cell $A$ marked in Figure 1a. A less local embodiment of this requirement is what Werfel and Nagpal call the *separation rule*. It states that two blocks can never be attached in the same row or column if all of the cells between them are ultimately meant to be occupied. For example, a block cannot be placed in the cell marked $B$ in Figure 1, since doing so would eventually force a block to be squeezed between two other adjacent blocks in that column. They introduce a distributed, perimeter following technique in which the robots trace counter-clockwise along the perimeter of the partially built structure until they reach the beginning of an incomplete row. They then move to the last block positioned in the row and place their block just before it. Their technique ensures that at all intermediate stages the partially built structure is a single, connected piece. In [2] they generalize this to building a restricted class of 3D shapes without holes.

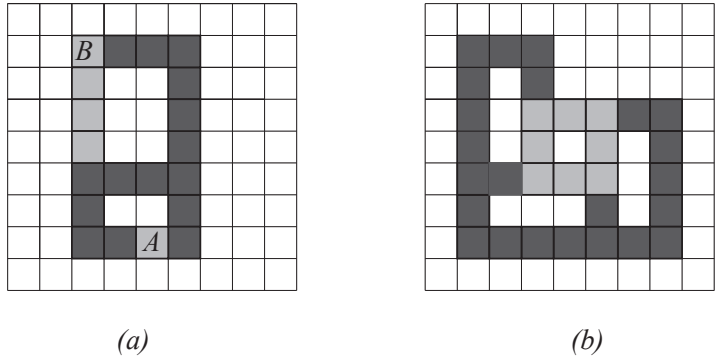

**Fig. 1.** Dark and lightly shaded cells represent the structure to be built. Blocks have already been placed in dark cells. (a) A block cannot be placed in cell $A$ because the gap is too narrow with the adjacent cells on the left and right both already occupied. A block cannot be placed in cell $B$ because it violates the separation rule. (b) Building has reached a dead-end since placing a block on any of the remaining adjacent lightly shaded sites violates the separation rule.

We consider the problem of building 2D block structures with holes, subject to the same two physical constraints imposed by Werfel and Nagpal's perimeter following approach: (i) no block can be placed between two other adjacent blocks on the same row or column (or equivalently, the separation rule is followed), and (ii) at all intermediate stages of construction the partially built structure is a single, connected piece. Our focus here is on determining a partial ordering on the blocks such that if the blocks are placed in accordance with this ordering, then the building process is able to complete without violating these two constraints. A partial ordering rather than a strict ordering is desirable since it allows for the possibility of concurrently attaching blocks during construction. For structures with holes, an arbitrary placing of the blocks according to constraints (i) and (ii) can lead to dead-end situations where it is impossible to complete the structure. See for example Figure 1b where the dark and light gray cells together represent the desired structure, with the dark cells representing the portion of the structure built so far. Observe that in each step of the construction so far the two constraints have been satisfied, and yet it is now impossible to complete without violating the separation rule.

In Section 2 we describe how we model this as a graph orientation problem. We present an algorithm for finding a partial ordering in Section 3, and we describe in Section 4 how to implement it $O(n^2)$ time, where $n$ is the number of blocks in the structure. In Section 5, we present a distributed algorithm that uses the partial ordering for collective construction.

## 2 Modeling as a Graph Orientation Problem

We model the problem of finding a partial ordering of the blocks as a graph orientation problem. Let $B$ be a 2D (face) connected, $n$ block structure to be built. Consider the dual graph $G$ of $B$, where each block is a vertex, and two vertices are connected by an (undirected) edge if their corresponding blocks in $B$ share a face. $G$ is a connected, orthogonal graph with each vertex having an $(x, y)$ location in the plane (as determined by the center of its corresponding block) and each edge having unit length. Our goal is to assign directions to the edges of $G$ such that the resulting directed graph $D$ has the following properties:

(P1) Each vertex has indegree of at most two, and the up to two incoming edges are orthogonal to each other.
(P2) $D$ is acyclic.
(P3) $D$ contains a root vertex, $r$, from which all other vertices in $D$ can be reached by following directed edges.

Each directed edge $\overrightarrow{uv}$ in $D$ represents the constraint that block $u$ must be attached before block $v$. Since property P2 ensures $D$ is acyclic, its edges determine a partial ordering on the blocks. If $B$'s blocks are attached in an order consistent with this partial ordering, then property P1 ensures that each block is attached before both its neighbors in the same row or column are attached (and so the separation rule is enforced). Property P3 ensures that if construction starts with the root block $r$, then at all intermediate stages of construction $B$ is one connected piece.

### 2.1 Orienting $G$

Our algorithm for orienting $G$ first decomposes it into what we call an *orientable sequence* consisting of $n$ components $(g_1, g_2, \ldots, g_n)$ such that $G = g_1 \cup g_2 \cup \ldots \cup g_n$. Let $S_n$ be an $n$ component orientable sequence.

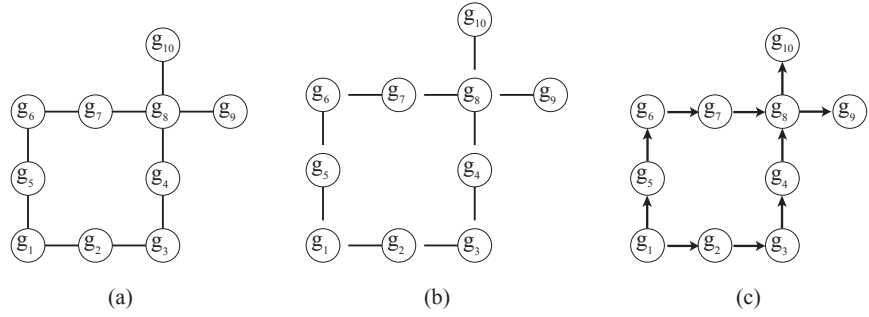

**Fig. 2.** (a) A graph $G$ decomposed into a ten component orientable sequence $(g_1, \ldots, g_{10})$. (b) $G$ shown blown apart into its individual components and their associated edges. Observe that $g_8$ is the only corner component. (c) Directed graph $D$.

We define $S_n$ recursively, since each subsequence $S_i = (g_1, \ldots, g_i)$ also defines an orientable sequence: $S_1 = (g_1)$ where $g_1$ is a single vertex. $S_i = (S_{i-1}, g_i)$ where $g_i$ is either a *terminal* or *corner* component. If $g_i$ is a terminal component, then it consists of a vertex $v_i$ and one adjacent edge whose other end point is a vertex in $S_{i-1}$. If $g_i$ is a corner component, then it consists of a vertex $v_i$ and two adjacent orthogonal edges whose other end points are in $S_{i-1}$. See Figure 2a,b. Observe that for each subsequence $S_i$, the graph formed by the union of its components, $g_1 \cup g_2 \cup \ldots \cup g_i$, is connected, for all $i = 1 \ldots n$.

We defer for now a discussion of how this decomposition is achieved, and instead show that given such a decomposition, we can oriented the edges of $G$ to obtain an oriented graph $D$ satisfying the three properties P1 - P3.

**Theorem 1.** *Let $G$ be a 2D orthogonal graph with $n$ vertices and unit length edges. If there exists an orientable sequence $(g_1, \ldots, g_n)$ such that $G = g_1 \cup g_2 \cup \ldots \cup g_n$, then it is possible to assign directions to the edges of $G$ such that the resulting directed graph $D$ has properties P1, P2, and P3 listed in Section 2.*

*Proof.* For each corner or terminal component $g_i$, let $d_i = g_i$ but with the edge(s) directed towards the component's vertex $v_i$. Let $D = \{d_1 \cup d_2 \cup \ldots \cup d_n\}$. See Figure 2c. $D$ clearly satisfies property (P1) since $d_1$ has indegree zero, terminal vertices have indegree one, and corner vertices have indegree two and the two incoming edges are orthogonal.

We show now by induction on $i$ that $D_i$ is connected with root $r = v_1$ (P3), and acyclic (P2). The base case is satisfied since $D_1 = \{d_1\}$ is the single vertex, $v_1$. Consider now $D_i$, for $i > 1$. $D_i$ is formed by adding component $d_i$ to $D_{i-1}$. By the inductive hypothesis, $v_1$ is the root of $D_{i-1}$, and so there exists a path from $v_1$ to every vertex in $D_{i-1}$, including the vertex (or the two vertices) to which $d_i$'s vertex, $v_i$, attaches. By following a directed edge to $v_i$, we get a path from $v_1$ to $v_i$ in $D_i$, thus establishing property P3. For property P2, it is easy to see that adding $d_i$ cannot create a cycle, since $v_i$ has no outgoing edges. □

## 3 Decomposing $G$ into an Orientable Sequence

Here we describe an algorithm for decomposing $G$ into an orientable sequence $S_n = (g_1, \ldots, g_n)$, as outlined in the pseudocode shown in Algorithms 1 and 2. Algorithm 1 forms the sequence in reverse order through $n - 1$ calls to the routine NEXTCOMPONENT, which is detailed in Algorithm 2. In Algorithm 1, observe that $G_n = G$, and in each subsequent iteration $G_i = G \setminus \{g_{i+1}, g_{i+2}, \ldots, g_n\}$, or in other words $G_i$ is the original graph minus the terminal and corner vertices of components $\{g_{i+1}, g_{i+2}, \ldots, g_n\}$ that have so far been identified.

The invocation NEXTCOMPONENT($G$) identifies a terminal or corner component, $g$, such that $G \setminus g$ is connected. Connectivity of $G \setminus g$ is necessary since we will use $g$ as the last component in an orientable sequence of $G$, and make additional calls to NEXTCOMPONENT to find an orientable sequence of $G \setminus g$. But if $G \setminus g$ is disconnected, then it has no orientable sequence. Therefore, in lines 3-6 of Algorithm 2, NEXTCOMPONENT finds a vertex $c$ that is either a degree one vertex or a degree two vertex with edges perpendicular, and then it checks if it is a non-cut vertex (line 7). If it is not a cut vertex, then $c$ and its one or two adjacent edges are returned as the component. If, however, $c$ is a cut vertex, then the search continues in the next iteration in one of the two parts of $G$ determined by removing $c$. Lines 10-20 set the appropriate variables in preparation for the next iteration.

For the purpose of explaining lines 10-20 of NEXTCOMPONENT, consider the graph $G$ illustrated in Figure 3. Observe that in the first iteration of the while loop, $c$ (selected in line 3) is the vertex labeled $c_1$ in Figure 3. We call such a vertex with maximal $x$ coordinate among all vertices with maximal $y$ coordinate the *northeast* vertex of the graph. Since this is the first iteration, $p = null$ and therefore the if statement in line 4 is not triggered. Observe that $c = c_1$ is a cut vertex, and therefore execution enters lines 10-20, which has the task of preparing for the next iteration in which searching for a component continues in one of the two subgraphs determined by $c$'s removal. Again because this is the first iteration and $p = null$, $G'$ is selected arbitrarily in line 12 to be one of the two subgraphs. Assume for this example that $G'$ is the subgraph containing the vertex labeled $p_1$. Observe that $p$ is set to $p_1$ in line 19, and in line 20 $G$ is set to be the smaller subgraph $G'$.

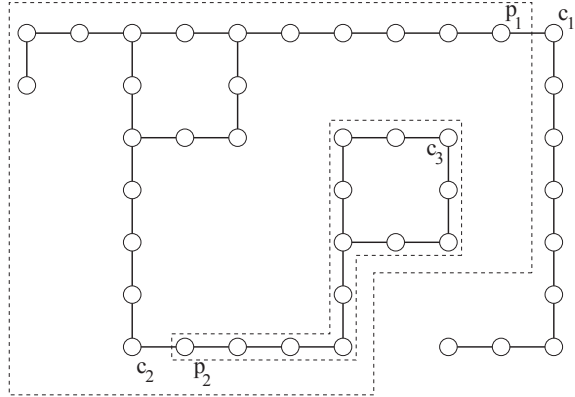

**Fig. 3.** Three iterations of NextComponent determine the next component of this graph to be the corner component with vertex labeled $c_3$. Candidates for a corner component found in iterations one and two involve vertices $c_1$ and $c_2$, but they are discarded because they are cut vertices. The dotted boundaries show the smaller graphs searched in iterations 2 and 3.

In the second iteration, the northeast vertex of the new (smaller) $G$ is $p_1$, but it is not a valid candidate for the next component because, although it appears to be a terminal vertex in $G$, it is not a terminal in the original graph since $c_1$ is also adjacent to it. This is detected in line 4, and so it then switches and takes the southwest vertex instead, which is the vertex marked $c_2$ in Figure 3. $c_2$ is a cut vertex, and therefore the search needs to continue in one of the two subgraphs determined by its removal. It is, however, important in this and subsequent iterations that the search continue in the subgraph $G'$ of $G$ not containing $p$ (line 14), since we don't want to contend with having multiple vertices in $G'$ having edges to vertices not in $G'$. Therefore, in the example, $G'$ is set to the subgraph not containing $p = p_1$, which is shown dotted in Figure 3. The point in $G'$ adjacent to $c_2$ is labeled $p_2$. In line 19, $p$ is set to $p_2$, in preparation for the next iteration. This continues in this manner until a non-cut vertex $c$ is found, which in the example occurs in the third iteration with the vertex $c_3$. At that point, $c_3$ and its two adjacent edges are returned as the next component.

## Algorithm 1 OrientableSequence(Graph $G$) : $\{g_1, g_2, \ldots, g_n\}$

$G_n \leftarrow G$
**for** i = n down to 2 **do**
    $g_i = \text{NextComponent}(G_i)$
    $G_{i-1} \leftarrow G_i \setminus g_i$
5: **end for**
$\{G_1$ consists of a single vertex$\}$ $g_1 \leftarrow G_1$

## Algorithm 2 NextComponent(Graph G):Component

**Ensure:** $G$ has at least 2 vertices
    Vertex $p \leftarrow null$
    **while** $true$ **do**
        Vertex $c \leftarrow$ the northeast vertex of $G$ (i.e., the vertex with maximum $x$ coordinate among those with maximum $y$ coordinate)
        **if** $c = p$ **then**
5:         $c \leftarrow$ the southwest vertex of $G$ (i.e., the vertex with minimum $x$ coordinate among those with minimum $y$ coordinate)
        **end if**
        **if** $c$ is not a cut vertex of $G$ **then**
            **return** the component consisting of $c$ and its (at most two) adjacent edge(s)
        **else**
10:         $\{c$ has degree two and removing it splits $G$ into two disjoint subgraphs$\}$
            **if** $p = null$ **then**
                $G' \leftarrow$ either of the two disjoint subgraphs created by removing $c$
            **else**
                $G' \leftarrow$ the subgraph created by the removal of $c$ that does not contain $p$
15:         **end if**
            **if** $G'$ consists of a single vertex $v$ **then**
                **return** the component consisting of $v$ and the edge connecting it to $c$
            **end if**
            $p \leftarrow$ the vertex in $G'$ adjacent to $c$
20:         $G \leftarrow G'$
        **end if**
    **end while**

To assist in the running time analysis of NEXTCOMPONENT in the next section, we prove the following invariant is true in each iteration of the loop of Algorithm 2. With this lemma, we will not need to recompute the cut vertices of each smaller graph $G'$.

**Lemma 1.** *After execution of line 19 but before the execution of line 20, for any vertex $v \in G'$ such that $v \neq p$, $v$ is a cut vertex in $G'$ if and only if it is a cut vertex in $G$.*

*Proof.* Each vertex in $G'$ has the same set of adjacent edges and vertices that it has in $G$, with the exception of $p$ which is missing the edge $(p, c)$. To see this, suppose for the sake of contradiction that it was false, and there there existed an edge other than $(p, c)$ connecting a vertex of $G'$ to a vertex of $G \setminus G'$. Such an edge cannot connect to $c$ since then $c$ would have degree larger than two in $G$, which is impossible since we selected $c$ as the northeast or southwest vertex and so it has degree at most two. Such an edge also cannot connect to any vertex in $G \setminus \{G', c\}$, since then $c$ is not a cut vertex of $G$. □

## 4 Running Time Analysis

In the proof of Lemma 2 below, we describe how to implement NEXTCOMPONENT in $O(n)$ time with presorting of the vertices.

**Lemma 2.** NEXTCOMPONENT*( $G$ ) can be implemented in $O(n)$ time if the vertices are presorted, where $n$ is the number of vertices in $G$.*

*Proof.* We begin by noting that since $G$ has maximum degree of four, the total size of $G$ (vertices and edges) is $O(n)$. We assume each vertex has associated fields called CUTVERTEX, UNDERCONSIDERATION, PRENUM, and DESCENDANTS. Before entering the loop of NEXTCOMPONENT, we identify the cut vertices of $G$ using a standard linear time, depth-first search algorithm such as described in [11] (pages 296-298), and we set their CUTVERTEX fields to true. Determining if $c$ is a non-cut vertex in line 7 is then just a matter of examining this field, since by Lemma 1 a vertex $c \neq p$ is a cut vertex in the current graph $G$ iff it is a cut vertex in the original $G$. The only exception is vertex $p$, which by the design of lines 4-6 is guaranteed not to be selected as $c$. Since there are fewer than $n$ loop iterations, the overall time spent checking if $c$ is a non-cut vertex in line 7 is $O(n)$.

The depth-first search used to identify the cut vertices also defines a preorder visit number for each vertex (i.e., each vertex is assigned a number according to the order in which it is first visited during the search) which we store in the PreNum field. The depth-first search also defines a depth-first search spanning tree with each edge of $G$ classified as a tree edge or a back edge. We set the Descendants field of each vertex to be the consecutive range of preorder numbers of its descendants in the spanning tree. It is commonly known that the leaves of the depth-first search spanning tree are non-cut vertices. Thus any cut vertex of degree two (and in particular, vertex $c$ in line 10 of NextComponent) is an internal tree node of the spanning tree with one tree edge connecting it to its parent and one tree edge connecting it to its only child. Thus its descendant range includes exactly the preorder numbers of the vertices in one of the two components created by its removal. The one exception is when the root of the spanning tree is a degree two cut vertex, since in that case it has no parent. The two components created on its removal are the subtrees rooted at its two children. So for the root, we store only the range of preorder numbers of the first child's subtree. In line 14 of NextComponent, the Descendants field of $c$ is used to determine in $O(1)$ time which of the two subgraphs created on $c$'s removal contains vertex $p$ (by comparing $p$'s preorder number to $c$'s descendants range). Initializing the fields PreNum and Descendants can be done during the depth-first search used to determine cut vertices with no asymptotic increase in its running time.

The Underconsideration field of each vertex is initially set to true outside of the loop. This field indicates if a vertex is still part of the graph $G$ in which the search for a terminal or corner vertex continues, since in each iteration $G$ is smaller. When $c$ is determined to be a cut vertex and $G'$ is selected in lines 11-15, we traverse the subgraph $G \setminus G'$ and mark each vertex's Underconsideration field as false. The marked vertices represent the part of the graph that is discarded, and thus will be ignored in future iterations of the loop. Over all iterations this takes a total of $O(n)$ time since each vertex is marked false at most once.

The Underconsideration field is important when finding the northeast and southwest vertices. We assume the vertices are presorted (in OrientableSequence) by decreasing $y$ coordinate, using decreasing $x$ coordinate as the secondary key. To find, for example, the northeast vertex of $G$ in lines 3-6, we scan the sorted list in the direction of decreasing $y$, starting where the scan left off in the previous iteration of the loop, since the next northeast vertex cannot be higher than the ones found in previous iterations. During the scan, vertices with false Underconsideration fields are ignored since they are a part of the graph that was discarded in an earlier iteration. Thus, at most one pass is made through the sorted list and the total time spent identifying northeast (and similarly southwest) vertices in NextComponent is $O(n)$ time.

The following theorem follows immediately, since OrientableSequence($G$) makes $n$ calls to NextComponent($G$).

**Theorem 2.** OrientableSequence( $G$ ) *can be implemented in $O(n^2)$ time, where $n$ is the number of vertices in $G$.*

## 5  Collective Construction Using the Partial Ordering

We describe a decentralized construction algorithm that uses the partial ordering to build a desired structure. Our method of finding components gives us a total ordering on the vertices (and thus the blocks) such that component $g_i$ is on the perimeter of the partially built structure $G_i = g_1 \cup g_2 \cup \ldots g_{i-1}$, So a perimeter following approach similar to that of Werfel et al. is possible, if one is satisfied with only being able to place one block at a time. More parallelism can be achieved using the partial ordering determined by the directed graph $D$ rather than the total ordering, but then the perimeter following technique no longer works since the structure surrounding a hole may be put in place before the structure inside the hole is completed, and thus the surrounding structure blocks (perimeter) access to the interior of the hole.

Taking a different approach, we assume the robots can move or hover just above the blocks in the partially built structure, such as would be possible in a zero or low gravity environment or underwater. Each robot stores a copy of the directed graph $D$ which encodes both the structure to be built and the partial ordering on the blocks. In addition, each robot stores a copy of an undirected spanning tree $S$ of $D$ which will guide the robot to an empty cell in need of a block. We assume the root block is put into place first, and (as suggested by Werfel and Nagpal in [1]) that it emits a beacon that the robots can use to find it.

The operation of each robot begins by fetching a block from the block cache, which is located far enough away from the building site so as not to interfere with construction. The robot then locates the root block and begins moving (or hovering) over the tops of the blocks of the partially built structure, visiting the blocks in a depth-first search traversal order of $S$. If the next position $c$ that the robot needs to move to has no block and if the up to two adjacent blocks that must be in place before $c$'s block can be placed (as determined by the incoming edges in $D$) are already in place, then the robot attaches the block it carries at position $c$. Once its block is attached, it immediately leaves the structure to fetch a new block from the cache and then repeat the depth first search traversal starting from the root block. Otherwise, if the blocks corresponding to $c$'s incoming edges are not yet in place, then the robot skips the portion of the depth first search that explores the subtree rooted at $c$ and continues with the rest of the traversal.

To speed the construction process and avoid having robots traverse completely built subtrees, we assume each block has a bit of read/write memory that is set to 1 if the subtree of $S$ rooted at the block is completely built and it is set to 0 otherwise. The robots can read and write this bit of memory when they are hovering over the block. When a robot attaches a non-leaf block in $S$, it sets the bit to 0. If the block is a leaf block, then the robot sets the bit to 1 instead, to indicate that its subtree (consisting of just the leaf block itself) is completely built. When a robot determines that all children of a block $b$ have their bits set to 1, then the robot sets $b$'s bit to one also.

With multiple robots performing a depth-first search traversal over the partially built structure, some coordination is necessary to prevent "traffic jams." Suppose, for example, two robots are hovering over adjacent blocks and they each want to move over the other's block. In this case, one robot can pass the other by moving vertically above it and going over while the other moves underneath. This coordinated move requires some local communication between the two robots. Also, before moving over a block, a robot must lock the block it wishes to move over to avoid two robots attempting to move over the same block at the same time and colliding. This can be handled through a test and set bit of memory on the block.

## 6   Conclusions

For any (face) connected 2D block structure, we showed there exists a partial ordering on the blocks such that if the blocks are attached in accordance with this ordering, at all intermediate phases of construction the partially built structure is connected and no block is ever squeezed in-between two other previously attached blocks. We provided an $O(n^2)$ algorithm for finding such an ordering, and described a distributed algorithm that uses the partial ordering for collective construction. This work leaves many open problems, in particular those related to building three dimensional structures. We believe the partial ordering algorithm can be adapted to find partial orderings for three dimensional (multi-layered) block structures with holes, but two difficult issues remain. First, it is not clear how to ensure that structure inside a hole is built before the structure surrounding the hole is put in place, thus preventing access to the hole's interior. Second, the robots need a way of searching out a position where they can attach their block, since 3D analogies for perimeter following or traversing over the top of the structure are not obvious. In 2D, it is an open question if finding the partial ordering can be done in less than $O(n^2)$ time.